\documentclass[letterpaper,english,aps,prb,reprint,superscriptaddress,longbibliography]{revtex4-2}
\usepackage[T1]{fontenc}
\usepackage[utf8]{inputenc}
\usepackage{color}
\usepackage{amsmath}
\usepackage{amssymb}
\usepackage{mathrsfs}
\usepackage{mathtools}
\usepackage{bm}
\usepackage{graphicx}
\usepackage{extarrows}
\usepackage{amsthm}

\usepackage[most]{tcolorbox}
\newtcbox{\othermathbox}[1][]{nobeforeafter, math upper, tcbox raise base, 
          enhanced, rounded corners, colback=black!5, colframe=black,
          left=0.7em, top=0.4em, right=0.7em, bottom=0.5em}
\usepackage{empheq}

\definecolor{MyYellow}{RGB}{248,199,82}

\usepackage{enumerate}

\usepackage[colorlinks,allcolors=blue]{hyperref}

\definecolor{LightBrown}{RGB}{255,188,0}
\definecolor{MiddleBrown}{RGB}{199,146,0}
\definecolor{DarkBrown}{RGB}{143,104,0}
\definecolor{DarkerBrown}{RGB}{87,62,0}
\definecolor{Purple}{RGB}{255,0,188}

\newcommand{\be}{\begin{equation}}
\newcommand{\ee}{\end{equation}}
\renewcommand{\zeta}{w}

\makeatletter

\pdfpageheight\paperheight
\pdfpagewidth\paperwidth

\usepackage{times}

\makeatother

\begin{document}
\title{Interaction-driven Weyl superfluids from topologically trivial bands in an optical Raman lattice}
\author{Yong-Huan Wang}
\affiliation{Hefei National Research Center for Physical Sciences at the Microscale and School of Physical Sciences, \\
    University of Science and Technology of China, Hefei 230026, China }
\affiliation{Hefei National Laboratory, University of Science and Technology of China, Hefei 230088, China}
\affiliation{Shanghai Research Center for Quantum Science and CAS Center for Excellence in Quantum Information and Quantum Physics, University of Science and Technology of China, Shanghai 201315, China}
\author{Xin-Chi Zhou}
\thanks{xinchi@pks.mpg.de}
\affiliation{Max Planck Institute for the Physics of Complex Systems, N\"othnitzer Stra\ss e 38, 01187 Dresden, Germany}
\begin{abstract}

We propose an interaction-driven route to Weyl superfluidity from topologically trivial noninteracting bands in a three-dimensional optical Raman lattice. The scheme builds on a pairing mechanism arising from the interplay between a large Zeeman splitting and an attractive Hubbard interaction. 
The large Zeeman splitting renders the noninteracting bands topologically trivial and suppresses  spin-flip hopping. 
These otherwise off-resonant spin-flip processes are resonantly restored when the attractive Hubbard interaction compensates the energy mismatch, generating occupation-dependent spin-orbit coupling that coherently couples neighboring spin-down fermion pairs to onsite doublons.  
Combining this density-dependent in-plane spin-orbit coupling with density-dependent spin-conserving interlayer tunneling extends the pairing mechanism to three dimensions. A self-consistent mean-field treatment of the microscopic model yields Weyl superfluids with two, four, or eight nodes, together with fully gapped weak topological superfluids. 
We further propose an experimental implementation with ultracold ${}^{40}\mathrm{K}$ atoms. Our results provide a route to Weyl superfluidity enabled by interaction-assisted spin-orbit coupling, without requiring nontrivial single-particle band topology.
\end{abstract}
\maketitle

\section{Introduction}
Weyl nodes are isolated, linearly dispersing band crossings
in three-dimensional momentum space that carry topological charges and act as monopoles of Berry
curvature~\cite{volovik2003_universe,wan2011_pyrochlore,burkov2011weyl,armitage2018_review}. These topological properties underlie characteristic phenomena
in Weyl semimetals, including Fermi-arc surface
states~\cite{lv2015_taas,xu2015_weyl_discovery}
and the chiral anomaly~\cite{armitage2018_review,jia2016_weyl}. Weyl superfluids extend this to Bogoliubov quasiparticle
spectra, realizing three-dimensional topological
phases that can support topologically protected  Majorana arcs~\cite{schnyder2015_nodal_review,sau2012_majorana_arcs,yang2014_dirac_weyl_sc}. Such phases have been proposed in spin-orbit-coupled Fermi
gases~\cite{gong2011_weyl_superfluid,xu2014_anisotropic_weyl_ff,xu2015_structured_weyl},
dipolar gases with interaction-induced chiral
pairing~\cite{liu2015_dipolar_weyl_sf},
and the three-dimensional repulsive Hubbard
model~\cite{kim2026_feshbach_weyl}. In particular, one route to Weyl superfluidity exploits the interplay of single-particle spin-orbit coupling (SOC), Zeeman fields, and attractive $s$-wave pairing~\cite{gong2011_weyl_superfluid,xu2014_anisotropic_weyl_ff,xu2015_structured_weyl}.

Synthetic SOC provides a versatile route to engineering topological band structures in ultracold atomic gases by coupling atomic internal states to their motion through laser fields~\cite{liu2009_induced_soc,
lin2011_soc_bec,dalibard2011_gauge_review,wang2012_soc_fermi,cheuk2012_soc_fermi,
goldman2014_gauge_review,zhai2015_soc_review,huang2016_2d_soc_fermi,zhang2018_soc_review,zhang2018_topological_review}. In particular, optical Raman lattices combine periodic trapping potentials with spatially modulated Raman couplings~\cite{liu2013_edge_spins,liu2016_chiral_raman,
zhou2017_spt_fermions,sun2018_quench_topology,song2018_spt_band,yi2019_topological_charges,ziegler2022_correlated_chern,
zhang2022_raman_superlattices,li2023_dissipative_raman,
zhang2023_floquet_bands}, enabling the realization of high-dimensional SOC~\cite{liu2014_2d_soc,liu2014_erratum,wu2016_optical_raman_lattice,
wang2018_dirac_rashba_weyl,sun2018_robust_2d_soc,song2019_nodal_line,lu2020_ideal_weyl,liang2023_qwz_fermions,zhou2023_dynamical_gauge,lin2026_unpaired_weyl}. This approach has led to the experimental realization of two-dimensional SOC~\cite{liu2014_2d_soc,liu2014_erratum,wu2016_optical_raman_lattice,sun2018_robust_2d_soc,liang2023_qwz_fermions} and an ideal Weyl-semimetal band with three-dimensional SOC~\cite{lu2020_ideal_weyl,wang2021_ideal_weyl_experiment}. These advances have motivated a variety of theoretical proposals for realizing topological superfluidity~\cite{liu2014_2d_soc,jia2019_topological_superfluids,poon2018_chiral_ff,gong2012_finite_temperature,qu2013_topological_ff,
zhang2013_topological_fflo,cao2014_gapless_ff}. A central challenge, however, arises from the competition between the Zeeman field and spin-singlet pairing. In conventional SOC-based schemes~\cite{zhang2008_swave_chiral,sato2009_nonabelian_sf,
zhu2011_nonabelian_statistics}, the Zeeman field required to access the topological regime also polarizes the system and tends to suppress the pairing favored by attractive interactions~\cite{gong2011_weyl_superfluid,
zhou2011_trapped_tsf}. This competition makes the experimental realization of topological superfluidity, particularly its Weyl counterpart, challenging.

\begin{figure}[tb]
    \centering
    \includegraphics[width=\linewidth]{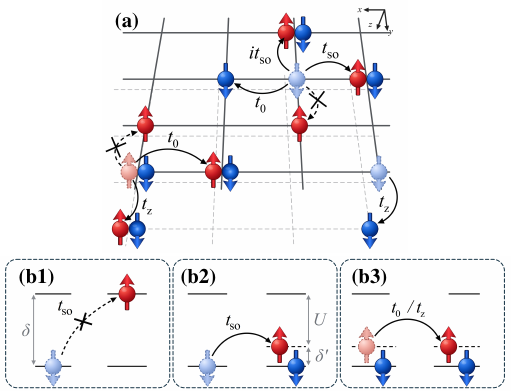}
    \caption{Density-dependent spin-orbit coupling in a three-dimensional optical
Raman lattice.
(a) Schematic of the lattice and allowed hopping processes in the
regime $U\simeq\delta\gg t_0,t_{\rm so},t_z$.
Spin-up fermions can only appear together with spin-down fermions on the same site.
(b1) Single-particle spin-flip hopping with amplitude $t_{\rm so}$
is suppressed by the large Zeeman splitting $\delta$.
(b2) The attractive interaction lowers the doublon energy by $U$,
bringing the transition
$\lvert\downarrow,\downarrow\rangle
\leftrightarrow\lvert0,\uparrow\downarrow\rangle$
into resonance when $\delta'=\delta-U\simeq0$.
(b3) Spin-conserving hopping of a spin-up fermion between sites
occupied by spin-down fermions, with amplitude $t_0$ in the
$xy$ plane and $t_z$ along the $z$ direction.}
    \label{fig:Fig1}
\end{figure}

An alternative route exploits an interaction-induced, density-dependent SOC proposed in a two-dimensional (2D) optical Raman lattice~\cite{zhou2023_dynamical_gauge}. In this scheme, a large Zeeman splitting suppresses the single-particle spin-flip hopping [Fig.~\ref{fig:Fig1}(b1)] and renders the noninteracting bands topologically trivial.  When a large attractive Hubbard interaction compensates the Zeeman energy mismatch, the otherwise off-resonant spin-flip processes become resonant for specific local particle configurations, allowing  two neighboring spin-down fermions to be converted into an onsite doublon and an empty site [Fig.~\ref{fig:Fig1}(b2)]. The resulting spin-flip hopping depends on the local particle configuration~\cite{zhou2023_dynamical_gauge,cheng2025_density_gauge}, giving rise to density-dependent SOC, or equivalently a non-Abelian dynamical gauge field. Importantly, these correlated processes also mediate an effective pair-hopping channel between onsite doublons and spin-triplet pairs of spin-down fermions occupying neighboring sites, supporting mobile pairs with mixed $s$- and $p$-wave pairing components. Thus, rather than competing, the Zeeman splitting and attractive interaction cooperate to generate both correlated SOC and the pairing mechanism responsible for topological superfluidity. The two-dimensional scheme was shown to support a fully gapped topological superfluid over broad ranges of interaction strength and particle filling. This naturally raises the question of whether the same mechanism can be extended to three dimensions to realize Weyl superfluidity without relying on topologically nontrivial single-particle bands.

In this work, we generalize the interaction-assisted pairing mechanism to three dimensions and establish, within a self-consistent mean-field framework, the emergence of Weyl superfluidity from topologically trivial noninteracting bands. We couple two-dimensional optical Raman lattices through spin-conserving interlayer tunneling, which is also constrained by local occupations in the projected low-energy manifold [Fig.~\ref{fig:Fig1}(a)]. 
Within the self-consistent mean-field description, the interlayer tunneling enters the Bogoliubov--de Gennes (BdG) Hamiltonian as a $k_z$-dependent shift of the effective chemical potential. 
Consequently, the three-dimensional system can be viewed as a family of two-dimensional momentum slices parametrized by $k_z$.

These 2D slices share the same self-consistently determined pairing field and densities, but can exhibit distinct topological properties as the effective chemical potential varies with $k_z$. To be specific, when all 2D momentum slices remain gapped with the same nonzero Chern number, the system is topologically equivalent to a stack of two-dimensional topological superfluids, realizing a fully gapped weak topological superfluid~\cite{teo2010_topological_defects,
hughes2014_weak_tsc,chiu2016_classification}. In contrast, the interlayer dispersion can drive the 2D momentum slices across topological transitions, where changes in the slice Chern number lead to bulk gap closings. 
Such gap closings form Weyl nodes when they are isolated and linearly dispersing. 

The resulting phase diagram features Weyl superfluids with two, four, or eight nodes, alongside fully gapped weak topological superfluids. We demonstrate that the interaction strength and particle filling control the nodal configurations and drive transitions between these phases. Finally, we outline an implementation of the proposed model using ultracold $^{40}\mathrm{K}$ atoms and provide estimates of the relevant experimental parameters. These results demonstrate how interaction-assisted pairing can generate Weyl topology in the Bogoliubov quasiparticle spectrum even when the underlying noninteracting bands are topologically trivial.

\begin{figure}[tb]
    \centering
    \includegraphics[width=\linewidth]
        {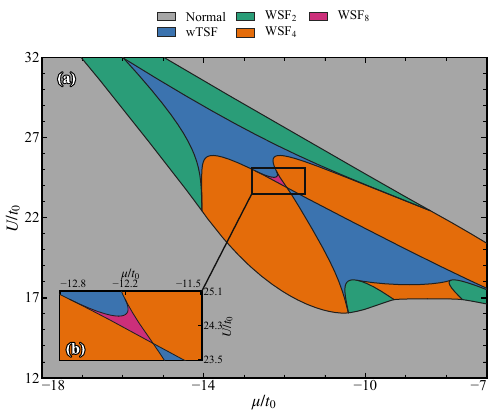}
    \caption{Mean-field phase diagram as a function of interaction strength $U$ and chemical potential $\mu$. (a) Normal, weak topological superfluid (wTSF), and Weyl
superfluid phases.
$\mathrm{WSF}_N$ denotes a Weyl superfluid with $N=2,4,8$
nodes in the three-dimensional Brillouin zone.
The wTSF is fully gapped, with the same nonzero Chern number
$\mathcal C(k_z)$ for all fixed-$k_z$ slices.
(b) Enlarged view of the boxed region in (a), showing the narrow
$\mathrm{WSF}_8$ region near $U\simeq\delta$.
Parameters are $\delta/t_0=24$ and $t_{\rm so}=t_z=t_0$.}
    \label{fig:phase_diagram}
\end{figure}

\begin{figure*}[t]
    \centering
    \includegraphics[width=0.99\textwidth]{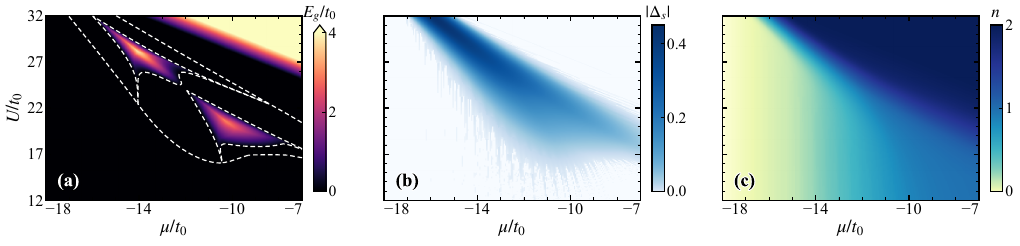}
    \caption{
        Self-consistent mean-field results.
(a) Bulk gap $E_g$, 
(b) onsite pairing amplitude $|\Delta_s|=|\langle c_{i\uparrow}c_{i\downarrow}\rangle|$, 
(c) and the density 
$n=\bar n_\uparrow+\bar n_\downarrow$ under different $U$ and $\mu$. 
White dashed lines in (a) outline the Weyl superfluid and
weak topological superfluid (wTSF) regions. Parameters are $\delta/t_0=24$ and $t_{\rm so}=t_z=t_0$.
    }
    \label{fig:mean-field-observables}
\end{figure*}

\section{Scheme}
We consider a three-dimensional lattice model with planar SOC, in the presence of a strong Zeeman field and an attractive on-site $s$-wave interaction, which is written as
\begin{equation}
H = H_0 + H_{\mathrm{int}},
\label{eq:lattice-hamiltonian}
\end{equation}
where the noninteracting part is
\begin{equation*}
H_0 =
\sum_{\vec{i},\,\nu=x,y,z}
(
\hat{\boldsymbol c}_{\vec{i}}^\dagger
T_\nu
\hat{\boldsymbol c}_{\vec{i}+\hat{\nu}}
+\mathrm{H.c.}
)
+
\sum_{\vec{i}}
\hat{\boldsymbol c}_{\vec{i}}^\dagger
(
-\mu \sigma_0
+\frac{\delta}{2}\sigma_z
)
\hat{\boldsymbol c}_{\vec{i}}.
\end{equation*}
Here, $\hat{\boldsymbol c}_{\vec{i}}
=(c_{\vec{i}\uparrow},c_{\vec{i}\downarrow})^{\top}$ is the spinor annihilation operator at the lattice site $\vec{i}=(i_x,i_y,i_z)$. The matrices $\sigma_{x,y,z}$ are the Pauli matrices acting in spin space, and $\sigma_0$ is the $2\times2$ identity matrix. 
The hopping matrices are $T_x = -t_0 \sigma_z + i t_{\mathrm{so}}\sigma_y$, $T_y = -t_0 \sigma_z + i t_{\mathrm{so}}\sigma_x$ and $T_z = -t_z \sigma_0$, respectively. 
Here, $t_0$ and $t_{\mathrm{so}}$ denote the spin-conserving and spin-flip hopping amplitudes in the $x$-$y$ plane, respectively, while $t_z$ denotes the spin-conserving hopping along the $z$ direction. The parameter $\mu$ is the chemical potential, and $\delta$ denotes the strength of the Zeeman splitting between the two spin states. The attractive Hubbard interaction is
\begin{equation*}
H_{\mathrm{int}}
=
-U\sum_{\vec{i}}
\hat n_{\vec{i}\uparrow}
\hat n_{\vec{i}\downarrow},
\quad U>0,
\end{equation*}
where $\hat n_{\vec{i}\sigma}
=c_{\vec{i}\sigma}^{\dagger}c_{\vec{i}\sigma}$ is the local particle-number operator for spin $\sigma\in\{ \uparrow,\downarrow\}$. The noninteracting bands are topologically trivial in the parameter regime considered in this work. To see this, we introduce the  momentum $\vec{k}=(k_x,k_y,k_z)$, with $\vec{k}_\perp=(k_x,k_y)$ denoting its in-plane component. The spin-dependent mass $m_0(\vec{k}_\perp)=\delta/2-2t_0(\cos k_x+\cos k_y)$ remains positive throughout the Brillouin zone for $\delta>8t_0$, ensuring a vanishing Chern number for every $k_z$ momentum slice. Since the interlayer tunneling contributes only a spin-independent dispersion, it does not alter the single-particle band topology. 

In this work, we focus on the density-dependent spin-orbit coupling regime, realized when the attractive Hubbard interaction and Zeeman splitting are the dominant energy scales, $U,\delta\gg t_0,t_{\mathrm{so}},t_z$, and nearly compensate each other, $U\simeq\delta$. In this regime, a bare single-particle spin flip is energetically forbidden because it creates a singly occupied $\lvert\uparrow\rangle$ state at the cost of the large Zeeman splitting [Fig.~\ref{fig:Fig1}(b1)].  The two-particle spin-flip process is, however, assisted by the attractive interaction, which compensates this Zeeman splitting, with the effective detuning denoted as $\delta'=\delta-U$ [Fig.~\ref{fig:Fig1}(b2)]. As a result, both spin-conserved and spin-flipped processes become density-dependent, in the sense that only hopping processes that do not create additional singly occupied spin-up states are allowed.

This two-site resonance has a direct many-body formulation. The large energy penalties imposed by $U$ and $\delta$ render the number of singly occupied spin-up sites, $\mathcal{N}_{\uparrow}\equiv\sum_{\vec{i}}\hat n_{\vec{i}\uparrow}(1-\hat n_{\vec{i}\downarrow})$, an emergent conserved quantity. In the limit $U,\delta\rightarrow\infty$ at fixed $\delta'=\delta-U$, sectors with different $\mathcal{N}_{\uparrow}$ decouple, and this conservation becomes exact. Restricting to the manifold relevant to the ground state, $\mathcal{N}_{\uparrow}=0$, the hopping part of Eq.~\eqref{eq:lattice-hamiltonian} reduces, to the lowest order, to
\begin{equation*}
H_{\mathrm{hop}}^{\mathrm{eff}}
=
\mathcal P_0
\sum_{\vec{i},\nu=x,y,z}
\left(
\hat{\boldsymbol c}_{\vec{i}}^{\dagger}
T_\nu
\hat{\boldsymbol c}_{\vec{i}+\hat\nu}
+\mathrm{H.c.}
\right)
\mathcal P_0,
\end{equation*}
where $\mathcal P_0=\prod_{\vec i}[1-\hat n_{\vec i\uparrow}(1-\hat n_{\vec i\downarrow})]$ projects onto the $\mathcal{N}_{\uparrow}=0$ sector. This projected form makes the occupation dependence of the hopping processes explicit: a spin-up fermion can hop only between sites occupied by spin-down fermions, a spin-down fermion can hop only without leaving or encountering a singly occupied spin-up site, and a spin flip is retained only as the resonant conversion $\lvert\downarrow,\downarrow\rangle\leftrightarrow\lvert\uparrow\downarrow,0\rangle$. Together with the direction-dependent Raman phases encoded in $T_x$ and $T_y$, these constraints generate a two-dimensional Dirac-type density-dependent SOC as shown in Fig.~\ref{fig:Fig1}.

The correlated spin-flip hopping also provides a direct pairing channel: the down-down state and the doublon state are coherently coupled by spin-flip hopping, enabling the resulting bound pair to propagate in the two-dimensional lattice. Schematically, two successive spin-flip
hopping processes along potentially different directions
$\nu,\nu'=x,y$ give
\begin{equation*}
\begin{aligned}
\left|
\downarrow_{\vec i},
\downarrow_{\vec i+\hat\nu},
0_{\vec i+\hat\nu+\hat\nu'},
\cdots\right\rangle
&\xleftrightarrow{\;t_{\rm so}^{\nu}\;}
\left|
0_{\vec i},
(\uparrow\downarrow)_{\vec i+\hat\nu},
0_{\vec i+\hat\nu+\hat\nu'},
\cdots\right\rangle
\\[-2pt]
&\xleftrightarrow{\;t_{\rm so}^{\nu'}\;}
\left|
0_{\vec i},
\downarrow_{\vec i+\hat\nu},
\downarrow_{\vec i+\hat\nu+\hat\nu'},
\cdots\right\rangle,
\end{aligned}
\end{equation*}
where $t_{\rm so}^{\nu}$ and $t_{\rm so}^{\nu'}$
represent the spin-flip hopping amplitudes along the two
in-plane directions, respectively. This is a microscopic mechanism for mobile pairs with mixed singlet $s$-wave and triplet $p$-wave components. 
In two dimensions, the resulting phase diagram contains fully gapped topological superfluids with Chern number $\mathcal{C}=\pm1$ over broad ranges of interaction and filling~\cite{zhou2023_dynamical_gauge}. Extending the system to three dimensions through spin-conserving interlayer tunneling offers a pathway to Weyl superfluidity without requiring nontrivial single-particle band topology. In the following section, we investigate the resulting superfluid phases using a  self-consistent mean-field treatment of the original microscopic Hamiltonian, while the projected model serves to elucidate the underlying occupation-dependent hopping mechanism.

\section{Phase diagram}

\subsection{Mean-field formulation}
We next study the self-consistent phase diagram of the system.
Within the mean-field treatment, we introduce 
the spin-resolved on-site densities $\bar n_{\uparrow}
=
\frac{1}{N_0}\sum_{\vec i}
\langle \hat n_{\vec i\uparrow}\rangle$ and $\bar n_{\downarrow}
=
\frac{1}{N_0}\sum_{\vec i}
\langle \hat n_{\vec i\downarrow}\rangle$ and the on-site $s$-wave pairing
order parameter $\Delta_s
=
\frac{1}{N_0}\sum_{\vec i}
\langle
c_{\vec i\uparrow}c_{\vec i\downarrow}
\rangle$, 
where $N_0$ denotes the total number of lattice sites.
In the Nambu basis $\Psi_{\vec k}
=
(
c_{\vec k\uparrow},
c_{\vec k\downarrow},
c^\dagger_{-\vec k\downarrow},
c^\dagger_{-\vec k\uparrow}
)^\top$, the mean-field Hamiltonian reads
\begin{equation}
H_{\mathrm{MF}}
=
\frac{1}{2}\sum_{\vec k}
\Psi^\dagger_{\vec k}
\mathcal H_{\mathrm{BdG}}(\vec k)
\Psi_{\vec k}
+E_{\mathrm{MF}},
\end{equation}
{where $E_{\mathrm{MF}}$ collects the constant terms, and}
\begin{equation}
\mathcal H_{\mathrm{BdG}}(\vec k)
=
\begin{pmatrix}
A_{\vec k} & B_{\vec k} & \Delta & 0\\
B_{\vec k}^{*} & C_{\vec k} & 0 & -\Delta\\
\Delta^{*} & 0 & -C_{-\vec k} & -B_{-\vec k}\\
0 & -\Delta^{*} & -B_{-\vec k}^{*} & -A_{-\vec k}
\end{pmatrix}.
\end{equation}
{Here $\Delta=U\Delta_s$ is the pairing term and}
\begin{equation}
\begin{aligned}
A_{\vec k}
&=M(\vec k_\perp)-\xi(k_z),\\
C_{\vec k}
&=-M(\vec k_\perp)-\xi(k_z),\\
B_{\vec k}
&=-2t_{\mathrm{so}}\sin k_y
+2it_{\mathrm{so}}\sin k_x,
\end{aligned}
\end{equation}
{with $M(\vec k_\perp)$ and the effective $k_z$-modulated chemical potential $\xi(k_z)$ given by}
\begin{equation}
\begin{aligned}
M(\vec k_\perp)
&=
\frac{\delta}{2}
-2t_0(\cos k_x+\cos k_y)
+Un_p,\\
\xi(k_z)
&=
\mu+U\bar n+2t_z\cos k_z.
\end{aligned}
\end{equation}
{Here 
$\bar n=(\bar n_\uparrow+\bar n_\downarrow)/2$,
and $n_p=(\bar n_\uparrow-\bar n_\downarrow)/2$}. 
Although the projector $\mathcal P_0$ is not explicitly enforced
in our mean-field treatment, the self-consistent solutions strongly
suppress the excluded singly occupied spin-up configurations,
whose average occupation
$\langle \hat n_{\vec i\uparrow}(1-\hat n_{\vec i\downarrow})\rangle$
remains below $6.9\times10^{-3}$ throughout the phase diagram
(see Appendix~\ref{app:single-occupation}).

\begin{figure*}[th]
    \centering
    \includegraphics[width=0.99\textwidth]{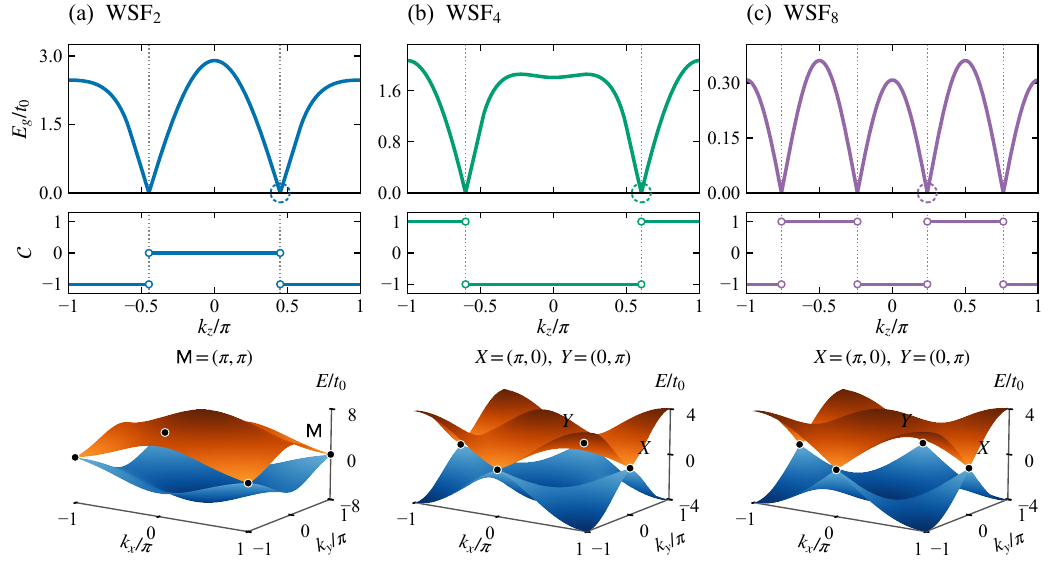}
    \caption{Topological characterization of Weyl superfluids.
Columns (a)--(c) show representative $\mathrm{WSF}_2$,
$\mathrm{WSF}_4$, and $\mathrm{WSF}_8$ states at
$(\mu/t_0,U/t_0)=(-14.6,30.5)$, $(-10.8,23.7)$, and
$(-12.1,24.2)$, respectively.
Top row: bulk gap $E_g(k_z)$ at a given $k_z$. Middle row: Chern number $\mathcal C(k_z)$ of the two occupied
Bogoliubov--de Gennes bands.
Vertical dotted lines mark gap-closing momenta;
the Chern-number jumps have magnitude one in (a) and two in (b), (c).
Bottom row: dispersions of the two middle bands,
$E_2$ (blue) and $E_3$ (orange), at
$k_z/\pi\simeq0.45$, $0.6$, and $0.24$
in (a)--(c), respectively.
The selected $k_z$ values are marked by dashed circles in the top row.
Black dots indicate Weyl nodes at $\mathsf M=(\pi,\pi)$ in (a)
and at $X=(\pi,0)$ and $Y=(0,\pi)$ in (b) and (c).
After identifying periodic boundary images, each displayed slice
contains one node in (a) and two in (b), (c).
Parameters are $\delta/t_0=24$ and $t_{\rm so}=t_z=t_0$.}
    \label{fig:weyl-signatures-multipoint}
\end{figure*}

\subsection{Phase diagram and momentum-slice topology}

 The $k_z$ dependence of the effective chemical potential $\xi(k_z)$ allows us to view the 3D system as a family of 2D BdG Hamiltonians parametrized by $k_z$. Each gapped momentum slice is characterized by a Chern number $\mathcal{C}(k_z)$. While a vanishing pairing order parameter $\Delta_s$ identifies the normal phase,  the topological properties of the superfluid phases with $\Delta_s\neq0$ are determined by the quasiparticle spectrum and $\mathcal{C}(k_z)$. When the spectrum remains fully gapped with a nonzero $\mathcal{C}(k_z)$ throughout the Brillouin zone, the system can be viewed as a stack of two-dimensional topological superfluids along the $z$ direction, realizing a weak topological superfluid (wTSF). In contrast, Weyl superfluidity arises when the gap closes at isolated momenta as $k_z$ varies, accompanied by changes in $\mathcal{C}(k_z)$. We denote a Weyl superfluid with $N$ Weyl nodes in the 3D Brillouin zone by $\mathrm{WSF}_N$.

The resulting phase diagram is shown in Fig.~\ref{fig:phase_diagram} for $\delta/t_0=24$ and $t_{\mathrm{so}}=t_z=t_0$. Remarkably, Weyl superfluidity extends over a broad range of interaction strengths $U$ and chemical potentials $\mu$. In addition to the normal phase and fully gapped weak topological superfluid phase, we identify three distinct Weyl superfluid phases $\mathrm{WSF}_N$ with $N=2,4,8$. While the $\mathrm{WSF}_2$ and $\mathrm{WSF}_4$ phases occupy extended regions, the $\mathrm{WSF}_8$ phase emerges within a narrow parameter regime near $U\simeq\delta$, as highlighted in Fig.~\ref{fig:phase_diagram}(b). 
As shown in Fig.~\ref{fig:mean-field-observables}(b), a finite pairing amplitude \(|\Delta_s|\) identifies the superfluid regions in Fig.~\ref{fig:phase_diagram}. Within these paired regions, the bulk BdG gap \(E_g\) in Fig.~\ref{fig:mean-field-observables}(a) remains finite in the wTSF phases but vanishes in the Weyl superfluid phases. {Figure~\ref{fig:mean-field-observables}(c) shows the corresponding
particle filling $n=\bar n_\uparrow+\bar n_\downarrow$, providing the filling conditions at which the different superfluid phases emerge for given model parameters.}

The distinction between fully gapped weak topological superfluids and gapless Weyl superfluids can be understood from the topology of two-dimensional momentum slices at different $k_z$. The interlayer dispersion enters the BdG Hamiltonian through the $k_z$-dependent effective chemical potential $\xi(k_z)$, allowing the three-dimensional BdG Hamiltonian to then be viewed as a family of two-dimensional systems with different chemical potentials. In the fully gapped superfluid regions of Fig.~\ref{fig:mean-field-observables}(a), the pairing amplitude remains finite, the quasiparticle gap stays open for all $k_z$, and the slice Chern number $\mathcal C(k_z)$ takes the constant values $-1$ and $+1$ in the upper and lower regions, respectively. These regions thus realize weak topological superfluids (wTSFs), topologically equivalent to stacks of two-dimensional topological superfluids along the $z$ direction (see Appendix~\ref{app:weak-topological-sf}). In contrast, in the Weyl superfluid phases, the variation of $\xi(k_z)$ drives the two-dimensional slices through topological transitions while the pairing amplitude remains finite. The quasiparticle gap consequently closes at isolated momenta, giving rise to Weyl nodes that separate slices with different Chern numbers. The locations of these nodes are determined below.

\subsection{Weyl-node counting and topological characterization}

The number and locations of Weyl nodes can be determined from the quasiparticle gap-closing conditions. For finite pairing $\Delta\neq0$, the quasiparticle gap can close only at momenta $\vec k^*=(\mathbf K,k_z^*)$, where $\mathbf K$ is one of the four in-plane high-symmetry momenta $\mathbf K=\Gamma=(0,0)$, $X=(\pi,0)$, $Y=(0,\pi)$, and $\mathsf M=(\pi,\pi)$, at which the spin-orbit coupling vanishes. The corresponding gap-closing condition is then given by 
\begin{equation}
    M_{\mathbf K}^2=\xi(k_z^*)^2+|\Delta|^2,
\end{equation}
where $M_{\mathbf K}=M(\mathbf K)$. As $k_z$ varies, the interlayer
dispersion modulates the effective chemical potential
$\xi(k_z)=\mu+U\bar n+2t_z\cos k_z$, which is symmetric under $k_z\rightarrow-k_z$.  Each admissible solution $0<k_z^*<\pi$ therefore gives rise to a pair of Weyl nodes at $(\mathbf K,\pm k_z^*)$  [Figs.~\ref{fig:weyl-signatures-multipoint}(a)--\ref{fig:weyl-signatures-multipoint}(c), top panels]. The number and positions of Weyl nodes are therefore determined by the admissible solutions at the different in-plane high-symmetry momenta. The analytical derivation and general node-counting rules are provided in Appendices~\ref{app:weyl-node-position} and~\ref{app:general_weyl_counting}, respectively.

The distinct Weyl superfluid phases can be understood
quantitatively by substituting the self-consistent mean-field
result into the analytical gap-closing conditions.
For the case $\delta/t_0=24$ and $t_z=t_0$, the effective masses at
the high-symmetry momenta are $M_\Gamma=8t_0+Un_p$,
$M_X=M_Y=12t_0+Un_p$, and $M_{\mathsf M}=16t_0+Un_p$.
We denote $\tilde\mu=\mu+U\bar n$ and
$R_{\mathbf K}=\sqrt{M_{\mathbf K}^2-|\Delta|^2}$,
the gap-closing condition yields two possible solutions,
\begin{equation}
\cos ( k_z^{*})
=\frac{-\tilde\mu\pm R_{\mathbf K}}{2t_z}.
\end{equation}
Each branch contributes a pair of Weyl nodes at
$(\mathbf K,\pm k_z^*)$, provided that
$M_{\mathbf K}^2>|\Delta|^2$ and
$|(-\tilde\mu\pm R_{\mathbf K})/(2t_z)|<1$.
The number and locations of Weyl nodes are thus determined
by the admissible solutions at the different high-symmetry
momenta $(\mathbf K,\pm k_z^*)$.

For the representative $\mathrm{WSF}_2$ state at $\mu=-14.6t_0$ and $U=30.5t_0$, the self-consistent
solution gives $\bar n\simeq 0.87$, $n_p\simeq -0.05$,
and $|\Delta_s|\simeq 0.26$, yielding
$\tilde\mu\simeq 12 t_0$ and
$R_{\mathsf M}\simeq 12.3 t_0$.
At $\mathsf M$ point $\mathbf K=(\pi,\pi)$, only one solution satisfies the gap-closing condition, with
$\cos k_z^*\simeq 0.16$.
At $\Gamma$ point $\mathbf K=(0,0)$, the condition $M_\Gamma^2>|\Delta|^2$
is not satisfied, and neither branch at $X$ and $Y$
admits a real solution for $k_z^*$.
Consequently, a single pair of Weyl nodes emerges at
$(\mathsf M,\pm k_z^*)$, with
$k_z^*/\pi\simeq 0.45$
[Fig.~\ref{fig:weyl-signatures-multipoint}(a)]. Since each gap closing involves a single Weyl node at a given $k_z$, the slice Chern number exhibits a unit jump $|\Delta\mathcal C|=1$.

Similarly, for the representative $\mathrm{WSF}_4$ state at $\mu=-10.8t_0$ and $U=23.7 t_0$, the self-consistent
solution gives $\bar n\simeq 0.71$, $n_p\simeq -0.22$,
and $|\Delta_s|\simeq 0.17$, yielding
$\tilde\mu\simeq 6.1 t_0$ and
$R_X=R_Y\simeq 5.5 t_0$.
Owing to the fourfold rotational symmetry, the
gap-closing conditions at $X$ and $Y$ are identical. 
At each of these high-symmetry momenta, one solution satisfies the gap-closing condition, giving
$\cos k_z^*\simeq -0.32$.
In contrast, at $\Gamma$ point, the mass term does not satisfy
$M_\Gamma^2>|\Delta|^2$, and the gap-closing equation at $\mathsf M$ point admits no real solutions for $k_z^*$. 
The simultaneous gap closings at $X$ and $Y$ points therefore
produce two pairs of Weyl nodes at
$k_z^*/\pi\simeq 0.6$, giving four Weyl nodes
in total [Fig.~\ref{fig:weyl-signatures-multipoint}(b)]. At each gap-closing $k_z$, the two Weyl nodes at $X$ and $Y$ carry identical topological charges, resulting in a Chern-number jump of magnitude $|\Delta\mathcal C|=2$. 

Finally, for the representative $\mathrm{WSF}_8$ state at $\mu=-12.1t_0$ and $U=24.2t_0$, the self-consistent solution gives $\bar n\simeq 0.5$, $n_p\simeq -0.25$, and $|\Delta_s|\simeq 0.24$. In this case, the effective chemical potential satisfies $\tilde\mu\simeq0$, while $R_X=R_Y\simeq 1.5\,t_0<2t_z$. The gap-closing condition therefore admits two distinct positive-$k_z$ solutions at both $X$ and $Y$, satisfying $\cos k_{z,1}^*=R_X/(2t_z)$ and $\cos k_{z,2}^*=-R_X/(2t_z)$, with $k_{z,1}^*/\pi\simeq 0.24$ and $k_{z,2}^*/\pi\simeq 0.76$. No additional nodes occur at $\Gamma$ or $\mathsf M$ points. Together with their negative-$k_z$ counterparts, these gap closings produce eight Weyl nodes in total [Fig.~\ref{fig:weyl-signatures-multipoint}(c)]. Importantly, the two distinct positive-$k_z$ solutions correspond to two successive topological transitions, each involving simultaneous gap closings at $X$ and $Y$. Since the two Weyl nodes at each transition carry identical topological charges, the slice Chern number exhibits two jumps of magnitude $|\Delta\mathcal C|=2$ for $0<k_z<\pi$. Their negative-$k_z$ counterparts give rise to two additional jumps with opposite signs, as shown in middle panel of Fig.~\ref{fig:weyl-signatures-multipoint}(c).

More generally, the analytical gap-closing conditions allow a wider variety of nodal configurations beyond the representative states discussed above. For example, $\mathrm{WSF}_2$ and $\mathrm{WSF}_4$ states can also host Weyl nodes at the in-plane momentum $\Gamma$ point. Furthermore, relaxing the choice $t_z=t_0$ can give rise to phases with different numbers of Weyl nodes beyond $N=2,4,8$, including a $\mathrm{WSF}_6$ phase. The number and locations of Weyl nodes can thus be controlled by the system parameters, as discussed in Appendix~\ref{app:general_weyl_counting}.

\begin{figure}[t!]
    \centering
    \includegraphics[width=\linewidth]{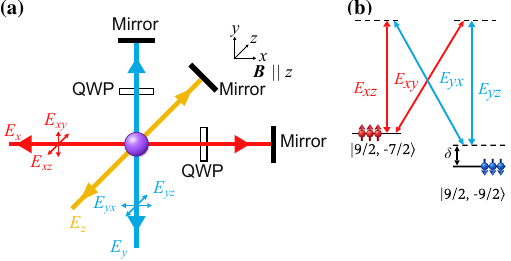}
    \caption{Experimental scheme. (a) Three retroreflected beams form the square optical Raman lattice in the $xy$ plane and the conventional lattice along $z$. The quarter-wave plates (QWPs) serve as phase retarders to generate the antisymmetric Raman coupling lattices. A bias magnetic field $\mathbf{B}$ applied along the $z$ direction generates the Zeeman splitting and defines the quantization axis. (b) Different polarization components of the laser beams generate Raman couplings between the ${}^{40}\mathrm{K}$ pseudospin states $\lvert 9/2,-9/2\rangle$ and $\lvert 9/2,-7/2\rangle$ through the indicated excited-state manifolds. The polarization components $E_{xy}$ and $E_{xz}$ (red), together with $E_{yx}$ and $E_{yz}$ (blue), generate the two Raman couplings. $\delta$ denotes the two-photon detuning.}
    \label{fig:experimental-scheme}
\end{figure}

\section{Experimental realization}
In this section, we 
propose a three-dimensional optical Raman lattice scheme to realize the predicted Weyl superfluid phases in ultracold fermions, building on established experimental techniques for two-dimensional spin-orbit coupling~\cite{liu2014_2d_soc,liu2014_erratum,wang2018_dirac_rashba_weyl,wu2016_optical_raman_lattice,sun2018_robust_2d_soc}. As illustrated in Fig.~\ref{fig:experimental-scheme}, two retroreflected laser beams in the $xy$ plane simultaneously generate a square optical lattice and two Raman couplings through their different polarization components. An additional retroreflected beam along the $z$ direction creates a conventional optical lattice, providing control over the spin-conserving interlayer tunneling. The resulting single-particle continuum Hamiltonian takes the form~\cite{liu2014_2d_soc,liu2014_erratum,sun2018_robust_2d_soc}
\begin{equation}
\begin{aligned}
H_{\mathrm{cont}}={}&
\left[\frac{\mathbf p^2}{2m}+V_{\mathrm{lat}}(\mathbf r)\right]\sigma_0
+\frac{\delta}{2}\sigma_z\\
&+M_1(x,y)\sigma_x-M_2(x,y)\sigma_y,
\end{aligned}
\label{eq:continuum-optical-hamiltonian}
\end{equation}
where $m$ is the atomic mass and $\mathbf p=-i\hbar\nabla$ and $\delta$ is the two-photon Raman detuning. The lattice potential is
\begin{equation}
\begin{aligned}
V_{\mathrm{lat}}(\mathbf r)={}&
V_{xy}\left[\sin^2(k_0x)+\sin^2(k_0y)\right]\\
&+V_z\sin^2(k_{L,z}z),
\end{aligned}
\label{eq:continuum-lattice-potential}
\end{equation}
and the Raman fields are given by 
\begin{equation}
\begin{aligned}
M_1(x,y)&=M_0\sin(k_0x)\cos(k_0y),\\
M_2(x,y)&=M_0\sin(k_0y)\cos(k_0x).
\end{aligned}
\label{eq:k40-raman-fields}
\end{equation}
Here, $V_{xy}$ and $V_z$ are the in-plane and axial lattice depths,
respectively, and $M_0$ is the common amplitude of the two Raman coupling
fields, expressed in energy units. The wave numbers are
$k_0=2\pi/\lambda_{xy}$ and $k_{L,z}=2\pi/\lambda_z$. The two Raman fields are antisymmetric about the lattice-site centers and have a relative phase of $\pi/2$, giving rise to two-dimensional spin-orbit coupling~\cite{wu2016_optical_raman_lattice,sun2018_robust_2d_soc}. In the tight-binding limit, the optical lattice and Raman fields generate spin-conserving hopping with amplitudes $t_0$ in the $xy$ plane and $t_z$ along the $z$ direction, together with spin-flip hopping amplitudes  $t_{\rm so}$ and $it_{\rm so}$ along the $x$ and $y$ directions, respectively. The two-photon detuning produces an effective Zeeman term $\delta/2$.

The optical Raman lattice offers flexible control over these microscopic parameters. The Raman detuning $\delta$ can be tuned by varying the frequency difference between the two in-plane beams, whereas $t_0$ and $t_{\rm so}$ can be controlled by adjusting the beam intensities and the relative fractions of their polarization components~\cite{sun2018_robust_2d_soc}. The axial lattice depth $V_z$ provides an additional control over $t_z$, allowing the interlayer dispersion to be tuned relative to the in-plane couplings.  The attractive interaction $U$ in Eq.~\eqref{eq:lattice-hamiltonian} can be tuned through a magnetic Feshbach resonance~\cite{loftus2002_feshbach,chin2010_feshbach}. The corresponding
resonance relation and tight-binding parameter estimates are presented in
Appendix~\ref{app:parameter-estimates}. In particular, when the large Zeeman splitting is compensated by the attractive Hubbard interaction, $U\simeq\delta\gg t_0,t_{\rm so},t_z$, the otherwise off-resonant spin-flip processes become resonant for specific local particle configurations, realizing the density-dependent spin-orbit coupling underlying our model.

As a concrete implementation, we consider ultracold ${}^{40}\mathrm{K}$ atoms with the pseudospin states $\lvert F=9/2,m_F=-9/2\rangle$ and $\lvert F=9/2,m_F=-7/2\rangle$. These states support a broad $s$-wave Feshbach resonance near $202.1\,\mathrm{G}$, with a width of several gauss, enabling continuous tuning of the attractive interaction~\cite{loftus2002_feshbach,chin2010_feshbach}. Raman-induced spin-orbit coupling has also been demonstrated in  continuum gases of ${}^{40}\mathrm{K}$ near this resonance, providing an experimental precedent for combining Raman spin coupling with resonantly enhanced interactions~\cite{fu2013_rf_soc_fermi}. The in-plane optical Raman lattice is formed using $764\,\mathrm{nm}$ laser beams~\cite{liu2014_2d_soc}, while a $1064\,\mathrm{nm}$ laser is used to generate the conventional lattice along the $z$ direction.

Representative parameter estimates  are $V_{xy}\simeq(5\text{--}9)E_r^{xy}$ and $V_z\simeq(3\text{--}6.5)E_r^z$, and $M_0\simeq(1.7\text{--}2.2)E_r^{xy}$, for which the three tunneling amplitudes can be comparable, $t_0/h\sim t_{\rm so}/h\sim t_z/h\simeq200\text{--}600\,\mathrm{Hz}$. Here, $E_r^{xy}=\hbar^2k_0^2/(2m)$ and
$E_r^z=\hbar^2k_{L,z}^2/(2m)$ are the in-plane
and axial recoil energies, respectively. Magnetic fields of approximately $203\text{--}205\,\mathrm{G}$ then yield $U\simeq(20\text{--}30)t_0$. Together with a Raman detuning satisfying $\delta\simeq U$ and an appropriate particle filling, these parameters provide access to the Weyl superfluid phases predicted by our self-consistent mean-field calculations. 
The corresponding parameter sets and details of their estimation are provided in Appendix~\ref{app:parameter-estimates} (Table~\ref{tab:k40-tight-binding-parameters}).

\section{Conclusions and discussion}
In summary, we have proposed a route to Weyl superfluidity in a three-dimensional optical Raman lattice based on density-dependent spin-orbit coupling. When a large attractive Hubbard interaction compensates a large Zeeman energy, otherwise off-resonant spin-flip hopping becomes resonant conditioned on local  configurations, giving rise to a non-Abelian dynamical gauge field.  Combining this interaction-assisted pairing mechanism with density-dependent spin-conserving interlayer tunneling, our self-consistent mean-field analysis identifies Weyl superfluids with two, four, or eight nodes, together with fully gapped weak topological superfluids.  

The topology of two-dimensional momentum slices provides a unified description of these phases. Weyl nodes emerge where the interlayer dispersion drives gap closings between slices with different Chern numbers, whereas the fully gapped weak topological phases retain a nonzero Chern number throughout the Brillouin zone. The Chern-number jumps across the nodes, together with their linear Bogoliubov dispersions, establish the Weyl character of the gapless phases. Tuning the interaction strength and particle filling controls the number of nodes and drives transitions between the different superfluids. We have also proposed an experimental protocol with ultracold $^{40}\mathrm{K}$ atoms in a three-dimensional optical Raman lattice  and provided estimates of the relevant parameters.

Our results are obtained within a self-consistent mean-field treatment. Nevertheless, the quantized chiral charges of the Weyl nodes protect the  Weyl superfluid phases against weak beyond-mean-field corrections. Beyond mean field, fluctuations of the pairing order parameter and residual quasiparticle interactions can renormalize the effective dispersion, thereby shifting the Weyl-node positions and phase boundaries. In our scheme, the large Hubbard attraction and Zeeman splitting primarily establish the resonant processes and local constraints underlying the density-dependent spin-orbit coupling. The stability of the resulting Weyl quasiparticles against additional interactions is therefore characterized by their residual low-energy couplings. For isolated, linearly dispersing Weyl nodes, the quasiparticle density of states vanishes as $\rho(E)\propto E^2$, and the coupling constants of local four-fermion interactions have a scaling dimension of $-2$, rendering weak short-range residual interactions perturbatively irrelevant~\cite{RoyGoswamiJuricic2017}. Consequently, weak residual interactions renormalize the Weyl cones without eliminating the topological charges.


\begin{acknowledgments}
We thank Xiong-Jun Liu, Long Zhang and Lin Zhang for their careful reading of the manuscript and valuable suggestions, and Shuai Chen, Jinyi Zhang, Yu-Min Hu, Meng Zeng, Sen Niu, Xiao-Dong Lin and Wei Jia for helpful discussions.

\textit{Use of AI tools.}
The authors used GPT-6 (OpenAI) to assist with the preparation  of this manuscript and the development of computational code.
The authors take full responsibility for the code, results, and scientific
content of this work.

\end{acknowledgments}

\section*{Data availability}

The numerical data supporting the findings of this study
are  available on Zenodo~\cite{Wang2026Data}.

\appendix

\section{Suppression of singly occupied spin-up states}
\label{app:single-occupation}
\begin{figure}[t!]
    \centering
    \includegraphics[width=\columnwidth]{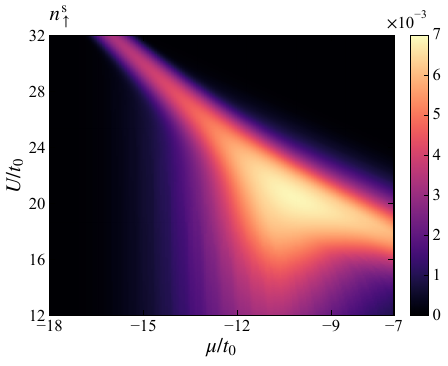}
    \caption{Mean-field occupation of singly occupied spin-up states 
    $n_\uparrow^{\mathrm s}$.
    Parameters are $\delta/t_0=24$ and $t_{\rm so}=t_z=t_0$.}
    \label{fig:single-up-occupation}
\end{figure}

In the main text, the self-consistent mean-field calculations are performed on the original Hamiltonian, without explicitly imposing the low-energy projection. Here, we numerically verify that the resulting mean-field solutions already exhibit a strong suppression of singly occupied spin-up states, consistent with the low-energy subspace underlying the effective model. To quantify this suppression, we evaluate the density of singly occupied spin-up states 
\begin{equation}
\begin{aligned}
n_\uparrow^{\mathrm s}
&=\langle\hat n_{i\uparrow}(1-\hat n_{i\downarrow})\rangle\\
&\simeq\bar n_\uparrow(1-\bar n_\downarrow)-|\Delta_s|^2.
\end{aligned}
\label{eq:app-single-up-mf}
\end{equation}
Here $\bar n_\uparrow$, $\bar n_\downarrow$, and the dimensionless
pairing amplitude $\Delta_s$ are obtained from the same
self-consistent solutions used in the main text.
As shown in Fig.~\ref{fig:single-up-occupation},
$n_\uparrow^{\mathrm s}<6.9\times10^{-3}$ throughout the 
parameter range, demonstrating that singly occupied spin-up states are strongly suppressed even without explicitly imposing the projection. This suppression originates from the interplay between the large Zeeman splitting and the strong attractive Hubbard interaction: the former energetically disfavors isolated spin-up fermions, while the latter favors doublon formation and enables resonant pair-hopping processes near $U\simeq\delta$.

Consequently, the self-consistent mean-field solutions predominantly occupy the low-energy configurations retained in the projected model. The negligible residual occupation thus provides a justification for the mean-field treatment of the original Hamiltonian and supports the effective interaction-assisted spin-orbit coupling mechanism underlying the Weyl superfluid phases.

\section{Weak topological superfluids}
\label{app:weak-topological-sf}

\begin{figure}[t!]
    \centering
    \includegraphics[width=\columnwidth]{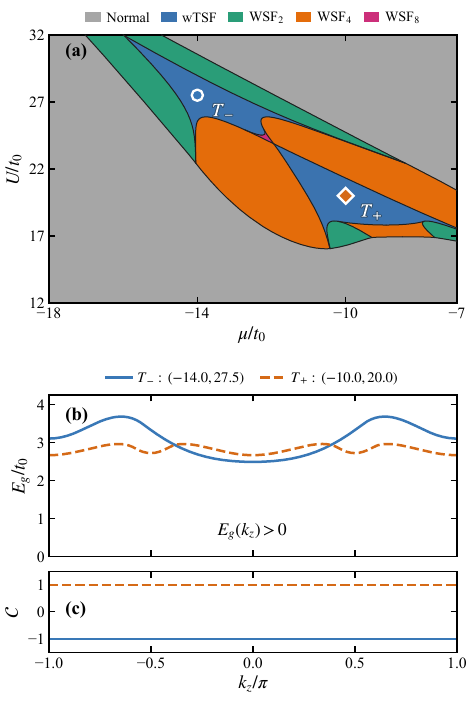}
    \caption{Topological characterization of fully gapped superfluids. 
    (a) Representative points $T_-$ (circle) and $T_+$ (diamond)
    in the phase diagram.
    (b) Quasiparticle gap under different $k_z$ and (c) total Chern number of the two occupied BdG bands
    for $T_-$ (blue solid) and $T_+$ (orange dashed).
    Both states are fully gapped, with $\mathcal{C}(k_z)=-1$ and $+1$,  respectively.
    Parameter pairs denote $(\mu/t_0,U/t_0)$;
    $\delta/t_0=24$ and $t_{\rm so}=t_z=t_0$.}
    \label{fig:weak-topological-sf}
\end{figure}

We verify the topological nature of two fully
gapped superfluids by showing their quasiparticle gaps
and slice Chern numbers. We choose two representative states, marked in Fig.~\ref{fig:weak-topological-sf}(a), at $T_-$:
$(\mu/t_0,U/t_0)=(-14.0,27.5)$ and $T_+$:
$(-10.0,20.0)$. 

As shown in Fig.~\ref{fig:weak-topological-sf}(b), the $k_z$-dependent quasiparticle gap
$E_g(k_z)$
remains finite throughout the Brillouin zone. Figure~\ref{fig:weak-topological-sf}(c) shows that the total Chern number of the two occupied BdG bands is independent of $k_z$ with  $\mathcal{C}(k_z)=-1$ for $T_-$ and
$\mathcal{C}(k_z)=+1$ for $T_+$ at every $k_z$.
The finite bulk gap associated with finite pairing and constant nonzero slice Chern number
identifies these states as three-dimensional weak topological
superfluids. The two points represent the upper and lower fully gapped regions, respectively. Since the bulk gap remains open within each region, the slice Chern number cannot change without a gap closing. The weak topological character therefore persists throughout the corresponding regions.

\section{Gap-closing conditions for Weyl nodes}
\label{app:weyl-node-position}


To explain the Weyl-node configurations found in the main text,
we derive the analytical gap-closing conditions of the BdG spectrum and determine the possible
node positions. We then examine the conditions for linear
dispersion near these gap-closing points.

In the Nambu basis
$\Psi_{\vec k}=(c_{\vec k\uparrow},c_{\vec k\downarrow},
c_{-\vec k\downarrow}^{\dagger},c_{-\vec k\uparrow}^{\dagger})^\top$,
the mean-field BdG Hamiltonian introduced in the main text reads
\begin{equation}
\mathcal H_{\rm BdG}(\vec k)=
\begin{pmatrix}
A_{\vec k} & B_{\vec k} & \Delta & 0\\
B_{\vec k}^* & C_{\vec k} & 0 & -\Delta\\
\Delta^* & 0 & -C_{-\vec k} & -B_{-\vec k}\\
0 & -\Delta^* & -B_{-\vec k}^* & -A_{-\vec k}
\end{pmatrix},
\label{eq:app-bdg-matrix}
\end{equation}
where $\Delta=U\Delta_s$ is the pairing potential, $U>0$ is the strength of the attractive interaction, and $\Delta_s$ is the
dimensionless pair amplitude. The
matrix elements are
\begin{equation}
\begin{aligned}
A_{\vec k}&=M(\vec k_\perp)-\xi(k_z),\\
C_{\vec k}&=-M(\vec k_\perp)-\xi(k_z),\\
B_{\vec k}&=h_x(k_y)-i h_y(k_x).
\end{aligned}
\label{eq:app-bdg-elements}
\end{equation}
Here $\vec k_\perp=(k_x,k_y)$, and the diagonal terms contain the mass term $M(\vec k_\perp)$
and effective chemical potential $\xi(k_z)$,
\begin{equation}
\begin{aligned}
M(\vec k_\perp)&=\frac{\delta}{2}
-2t_0(\cos k_x+\cos k_y)+Un_p,\\
\xi(k_z)&=\tilde\mu+2t_z\cos k_z.
\end{aligned}
\label{eq:app-M-xi}
\end{equation}
Here $\tilde\mu=\mu+U\bar n$ is the $k_z$-independent part of $\xi(k_z)$. The hopping amplitudes $t_0$ and $t_z$ describe spin-conserving
tunneling within and between the layers, respectively. $\delta$
is the Zeeman detuning and $\mu$ is the chemical potential.
The averaged density $\bar{n}$ and spin polarization $n_p$ are given by
\begin{equation}
\bar n=\frac{\bar n_\uparrow+\bar n_\downarrow}{2},\qquad
n_p=\frac{\bar n_\uparrow-\bar n_\downarrow}{2},
\label{eq:app-mean-field-scales}
\end{equation}
where $\bar n_\sigma$ is the density of spin-$\sigma$ fermions
per site. The off-diagonal spin matrix element $B_{\vec k}$ arises from the
in-plane SOC, whose two components are
\begin{equation}
h_x(k_y)=-2t_{\rm so}\sin k_y,\qquad
h_y(k_x)=-2t_{\rm so}\sin k_x,
\label{eq:app-soc-fields}
\end{equation}
where $t_{\rm so}$ is the spin-flip hopping amplitude. We choose a gauge in which $\Delta$ is real. To obtain the quasiparticle spectrum, we rewrite
Eq.~\eqref{eq:app-bdg-matrix} in the compact form
\begin{equation}
\mathcal H_{\rm BdG}(\vec k)=
h_x\sigma_x+h_y\sigma_y+M\sigma_z-\xi\tau_z
+\Delta\tau_x\sigma_z,
\label{eq:app-bdg-compact}
\end{equation}
where $\tau_i$ and $\sigma_i$ are Pauli matrices in the
particle--hole and spin sectors of the chosen Nambu basis,
respectively.

Particle--hole symmetry organizes the four eigenvalues as
\begin{equation}
E_{\lambda,\eta}(\vec{k})=\lambda E_\eta(\vec{k}),
\qquad \lambda=\pm1,\quad \eta=\pm1,
\label{eq:app-spectrum-pairs}
\end{equation}
where
\begin{equation}
\begin{aligned}
E_\eta^2={}&\xi^2+M^2+h_\perp^2+\Delta^2\\
&+2\eta\sqrt{\xi^2\left(M^2+h_\perp^2\right)+M^2\Delta^2},\\
h_\perp^2={}&h_x^2+h_y^2.
\end{aligned}
\label{eq:app-spectrum}
\end{equation}
The quasiparticle gap closes when the lower positive-energy
branch $E_-$ reaches zero. Using Eq.~\eqref{eq:app-spectrum},
we rewrite this condition as a sum of two nonnegative terms,
\begin{equation}
\left(\xi^2+\Delta^2-M^2-h_\perp^2\right)^2
+4\Delta^2h_\perp^2=0.
\label{eq:app-zero-sum}
\end{equation}
For a superfluid with \(\Delta\neq0\), the two terms must vanish
separately, so the gap-closing conditions are
\begin{equation}
h_x=h_y=0,\qquad M^2=\xi^2+\Delta^2.
\label{eq:app-gap-closing}
\end{equation}
From the SOC fields in Eq.~\eqref{eq:app-soc-fields}, the first condition
restricts the in-plane momentum to
\begin{equation}
\mathbf{K}\in\left\{\Gamma=(0,0),\ X=(\pi,0),\
Y=(0,\pi),\ \mathsf M=(\pi,\pi)\right\}.
\label{eq:app-high-symmetry-points}
\end{equation}
The corresponding Hartree-renormalized masses are
\begin{equation}
\begin{aligned}
M_\Gamma&=\frac{\delta}{2}-4t_0+Un_p,&
M_X=M_Y&=\frac{\delta}{2}+Un_p,\\
M_{\mathsf M}&=\frac{\delta}{2}+4t_0+Un_p.
\end{aligned}
\label{eq:app-high-symmetry-masses}
\end{equation}
At any such point the lower branch simplifies to
\begin{equation}
E_-(\mathbf{K},k_z)=
\left|\,|M_{\mathbf{K}}|-\sqrt{\xi(k_z)^2+\Delta^2}\,\right|.
\label{eq:app-high-symmetry-gap}
\end{equation}
The gap therefore closes at $(\mathbf K,k_z^*)$ when
\begin{equation}
M_{\mathbf{K}}^2=
\left(\tilde\mu+2t_z\cos k_z^*\right)^2+\Delta^2.
\label{eq:app-node-equation}
\end{equation}
With $R_{\mathbf K}\equiv\sqrt{M_{\mathbf K}^2-|\Delta|^2}$,
the gap-closing condition gives
\begin{equation}
\cos k_z^*=\frac{-\tilde\mu\pm R_{\mathbf K}}{2t_z}.
\label{eq:app-kz-solution}
\end{equation}
A physical solution exists only when
\begin{equation}
M_{\mathbf K}^2 \geq \lvert\Delta\rvert^2,
\qquad
\left\lvert\frac{-\tilde\mu\pm R_{\mathbf K}}{2t_z}\right\rvert \leq 1.
\label{eq:app-existence}
\end{equation}
For the representative $\mathrm{WSF}_4$ point, the valid solution occurs at both
\(X\) and \(Y\), producing the four nodes
\begin{equation}
(k_x,k_y,k_z)=
(\pi,0,\pm k_z^*),\qquad(0,\pi,\pm k_z^*).
\label{eq:app-wsf4-nodes}
\end{equation}
Their Weyl character follows from the linear variation of all three
independent coefficients.  For example, near
\((\pi,0,k_z^*)\), with \(\vec q=(k_x-\pi,k_y,k_z-k_z^*)\) and \(\xi^*\equiv\xi(k_z^*)\),
\begin{equation}
\begin{aligned}
h_y&\simeq2t_{\rm so}q_x,\qquad
h_x\simeq-2t_{\rm so}q_y,\\
\xi-\xi^*&\simeq-2t_z\sin k_z^*\,q_z.
\end{aligned}
\label{eq:app-linear-weyl}
\end{equation}
Hence the two touching middle bands disperse linearly in all three momentum
directions when \(t_{\rm so}t_z\xi^*\sin k_z^*\neq0\).

To check the analytical node positions against the numerical spectrum,
we use the self-consistent $\mathrm{WSF}_4$ state shown in
Fig.~\ref{fig:weyl-signatures-multipoint}(b), with parameters
\(U/t_0=23.7\), \(\mu/t_0=-10.8\), \(\delta/t_0=24\), and
\(t_z/t_0=1\), with
\(\bar n_\uparrow=0.495\), \(\bar n_\downarrow=0.931\), and
\(\Delta_s=0.173\).  In units of \(t_0\), these values give
\(\bar n=0.713\), \(n_p=-0.218\), \(\Delta=4.1\), and
\(M_X=M_Y=6.83\).  The only admissible branch of
Eq.~\eqref{eq:app-kz-solution} yields \(\cos k_z^*\simeq-0.317\), or
\(k_z^*/\pi\simeq0.603\). The predicted momenta $k_z=\pm k_z^*$ coincide with the gap closings
and Chern-number jumps in the top and middle panels of
Fig.~\ref{fig:weyl-signatures-multipoint}(b).

\section{General counting and locations of Weyl nodes}
\label{app:general_weyl_counting}

In this appendix, we establish a general counting rule for the Weyl
nodes of the present model using the gap-closing conditions derived in
Appendix~\ref{app:weyl-node-position}. We explain how the $\mathrm{WSF}_2$, $\mathrm{WSF}_4$, and $\mathrm{WSF}_8$ phases
presented in the main text fit into this counting scheme and discuss
additional nodal configurations beyond the parameter choice used in
the main-text phase diagram. 

For finite pairing $\Delta\neq0$, the gap-closing conditions derived in
Appendix~\ref{app:weyl-node-position} restrict the Weyl nodes to the four
in-plane high-symmetry momenta $\Gamma$, $X$, $Y$, and $\mathsf M$
[Eq.~\eqref{eq:app-high-symmetry-points}]. Let $\mathcal N_\Gamma$, $\mathcal N_X$, $\mathcal N_Y$,
and $\mathcal N_\mathsf M$ denote the respective numbers of Weyl nodes
over all $k_z$, counting periodically equivalent nodes only once. Since $M_X=M_Y$ [Eq.~\eqref{eq:app-high-symmetry-masses}], the $X$ and $Y$ lines have identical
node counts, $\mathcal N_X=\mathcal N_Y$. The total number of Weyl nodes
in the three-dimensional Brillouin zone is therefore
\begin{equation}
\begin{aligned}
\mathcal N
&=\mathcal N_\Gamma+\mathcal N_\mathsf M+\mathcal N_X+\mathcal N_Y\\
&=\mathcal N_\Gamma+\mathcal N_\mathsf M+2\mathcal N_X.
\end{aligned}
\label{eq:D_total_count}
\end{equation}
Here the factor of two accounts for the two symmetry-related lines $X$
and $Y$; the nodes at opposite $k_z$ are already included in each
contribution.

To determine each contribution, Eq.~\eqref{eq:app-kz-solution} provides two branches for
$\cos k_z^*$, distinguished by the sign $s=\pm1$. Assuming
$t_{\rm so},t_z,\Delta\neq0$, a branch produces ordinary Weyl nodes when
the inequalities in Eq.~\eqref{eq:app-existence} hold strictly:
\begin{equation}
\begin{gathered}
M_{\mathbf K}^2>|\Delta|^2,\\
\left\lvert\frac{-\tilde\mu+sR_{\mathbf K}}{2t_z}\right\rvert<1,
\qquad s=\pm1.
\end{gathered}
\label{eq:D_admissible_branch}
\end{equation}
Each admissible branch yields a pair of nodes at
$(\mathbf K,\pm k_z^*)$ with $0<k_z^*<\pi$. Thus, each high-symmetry momentum contributes 0, 2, or 4
Weyl nodes, depending on whether zero, one, or both
branches are admissible. In particular,
\begin{equation}
\begin{gathered}
\mathcal N_\Gamma,\mathcal N_\mathsf M,\mathcal N_X\in\{0,2,4\},\\
\mathcal N_X+\mathcal N_Y=2\mathcal N_X\in\{0,4,8\}.
\end{gathered}
\label{eq:D_line_counts}
\end{equation}
The strict inequalities exclude points where Weyl nodes merge.
When $M_{\mathbf K}^2=|\Delta|^2$, the two branches of
Eq.~\eqref{eq:app-kz-solution} give the same node positions,
and the corresponding nodes merge. When $|\cos k_z^*|=1$,
the nodes at $k_z=\pm k_z^*$ meet at $k_z=0$ or $\pi$.
In both cases, the dispersion along $k_z$ is no longer linear
[Eq.~\eqref{eq:app-linear-weyl}], so these merging points
are excluded from the Weyl-node count.

For $\delta/t_0=24$ and $t_{\rm so}=t_z=t_0$, the self-consistent phase
diagram presented in Fig.~\ref{fig:phase_diagram} contains the $\mathrm{WSF}_2$, $\mathrm{WSF}_4$, and $\mathrm{WSF}_8$
phases. The representative states in Fig.~\ref{fig:weyl-signatures-multipoint} have
\begin{equation}
(\mathcal N_\Gamma,\mathcal N_\mathsf M,\mathcal N_X)=
\begin{cases}
(0,2,0), & \mathrm{WSF}_2,\\
(0,0,2), & \mathrm{WSF}_4,\\
(0,0,4), & \mathrm{WSF}_8.
\end{cases}
\label{eq:D_main_text_examples}
\end{equation}
These configurations arise from one admissible branch at $\mathsf M$, one
admissible branch at each of $X$ and $Y$, and two admissible branches at
each of $X$ and $Y$, respectively. 

More generally, Weyl nodes can occur simultaneously on several
inequivalent high-symmetry lines, with their total number given by
Eq.~\eqref{eq:D_total_count}. Combining the contributions in
Eq.~\eqref{eq:D_line_counts} gives the possible total node counts
\begin{equation}
\mathcal N\in\{0,2,4,6,8,10,12,14,16\},
\label{eq:D_algebraic_counts}
\end{equation}
where $\mathcal N=0$ denotes the absence of Weyl nodes. The upper bound
$\mathcal N=16$ requires both branches to be admissible on all four
lines. These contributions are not independently adjustable: their
masses are related by Eq.~\eqref{eq:app-high-symmetry-masses}, and all lines share the same
self-consistent $\Delta$, $\bar n$, and $n_p$. The self-consistent solutions determine which of the node counts
in Eq.~\eqref{eq:D_algebraic_counts} are realized.

Varying the interlayer hopping $t_z$ beyond the choice $t_z=t_0$ used in the main text
allows additional Weyl-node configurations.
As $k_z$ varies, $\tilde\mu+2t_z\cos k_z$
spans an energy interval of width $4|t_z|$. Increasing $|t_z|$
broadens this interval and can allow gap closings on several
inequivalent high-symmetry lines. In particular, at $t_z/t_0=1.5$,
$t_{\rm so}/t_0=1$, $\delta/t_0=24$, $U/t_0=27$, and
$\mu/t_0=-14.2$, the self-consistent solution yields a $\mathrm{WSF}_6$ state.
Specifically, the $X$ and $Y$ lines each host a pair of
Weyl nodes at $k_z=\pm k_z^*$ with $k_z^*/\pi\simeq0.193$, contributing four
nodes in total, while the $\mathsf M$ line hosts another pair at $k_z=\pm k_z^*$ with $k_z^*/\pi\simeq0.804$.
The $\Gamma$ line remains gapped.
Thus, $(\mathcal N_\Gamma,\mathcal N_\mathsf M,\mathcal N_X)=(0,2,2)$,
giving $\mathcal N=0+2+2\times2=6$.

\section{Feshbach resonance and tight-binding parameter estimates}
\label{app:parameter-estimates}

\begin{table*}[tb]
\caption{Representative experimental parameter sets for $\lambda_{xy}=764\,\mathrm{nm}$ and
$\lambda_z=1064\,\mathrm{nm}$. The parameter sets give
values of $U/t_0$ close to 30 or 20. Hoppings are in Hz and $U/h$ is in kHz.}
\label{tab:k40-tight-binding-parameters}
\centering
\setlength{\tabcolsep}{3.4pt}
\renewcommand{\arraystretch}{0.94}
\begin{ruledtabular}
\begin{tabular}{ccccrrrrrr}
$V_{xy}/E_r^{xy}$ & $V_z/E_r^z$ & $M_0/E_r^{xy}$ & $B$ (G) &
$a_s/a_0$ & $t_0/h$ (Hz) & $t_{\rm so}/h$ (Hz) & $t_z/h$ (Hz) & $U/h$ (kHz) & $U/t_0$ \\
5.0 & 3.0 & 1.70 & 202.68 & $-2226$ & 563 & 567 & 490 & 16.99 & 30 \\
5.0 & 3.0 & 1.70 & 202.94 & $-1483$ & 563 & 567 & 490 & 11.32 & 20 \\
7.0 & 4.5 & 1.95 & 203.38 & $-914$  & 337 & 343 & 331 & 10.13 & 30 \\
7.0 & 4.5 & 1.95 & 203.88 & $-608$  & 337 & 343 & 331 &  6.74 & 20 \\
9.0 & 6.5 & 2.20 & 204.45 & $-418$  & 207 & 215 & 197 &  6.21 & 30 \\
9.0 & 6.5 & 2.20 & 205.17 & $-279$  & 207 & 215 & 197 &  4.15 & 20 \\
\end{tabular}
\end{ruledtabular}
\end{table*}

\begin{figure}[t!]
    \centering
    \includegraphics[width=0.98\columnwidth]{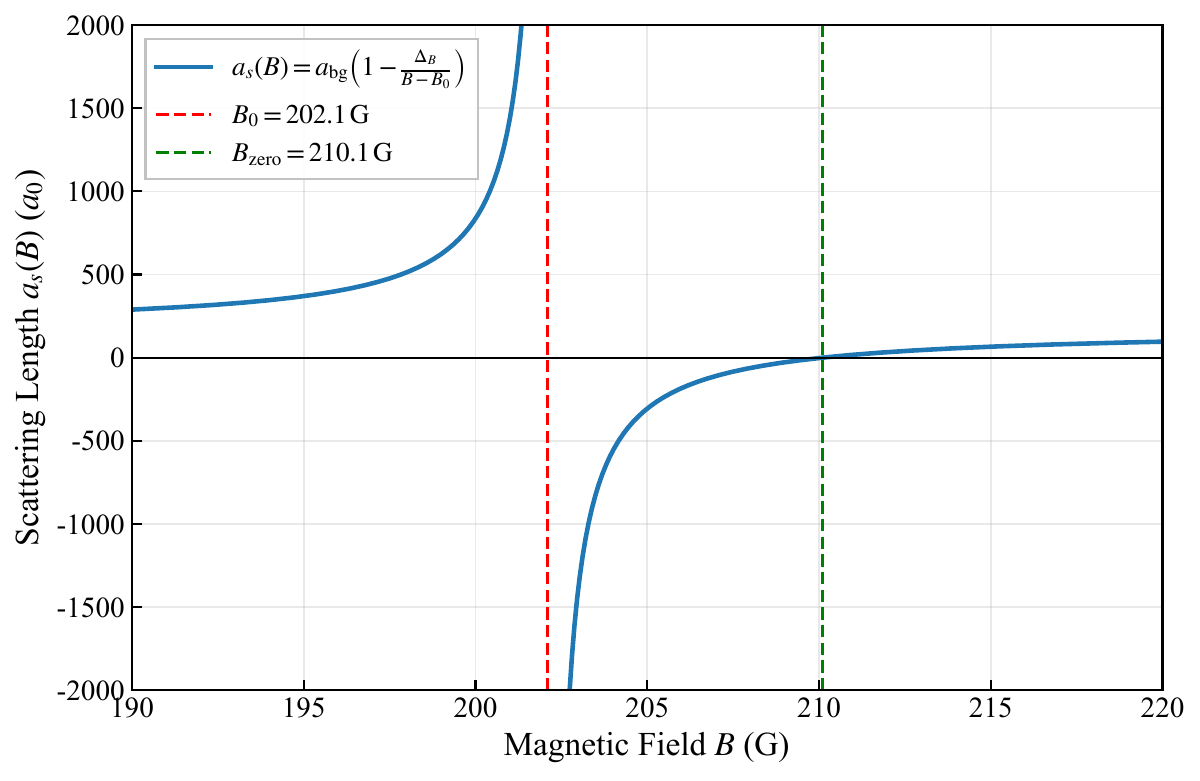}
   \caption{$s$-wave Feshbach resonance in the ${}^{40}\mathrm{K}$ mixture
    $\lvert9/2,-9/2\rangle$ and $\lvert9/2,-7/2\rangle$.
    The solid curve shows the scattering length as a function of magnetic field,
    and the dashed lines mark $B_0$ and $B_{\rm zero}$.}
    \label{fig:k40-feshbach-resonance}
\end{figure}

In this appendix, we estimate the hopping amplitudes and onsite interaction for the proposed optical Raman lattice. We use the lowest-band Wannier functions of the separable lattice potential in Eq.~\eqref{eq:continuum-lattice-potential} to relate the tight-binding parameters to the optical lattice depths, Raman coupling strength, and magnetic field. For the ${}^{40}\mathrm{K}$ mixture
$\lvert\downarrow\rangle=\lvert9/2,-9/2\rangle$ and
$\lvert\uparrow\rangle=\lvert9/2,-7/2\rangle$, the $s$-wave scattering length is
\begin{equation}
    a_s(B)=a_{\rm bg}\left(1-\frac{\Delta_B}{B-B_0}\right),
    \label{eq:k40-feshbach}
\end{equation}
with $B_0=202.1\,{\rm G}$, $\Delta_B=8.0\,{\rm G}$, and
$a_{\rm bg}=174a_0$~\cite{chin2010_feshbach}. As shown in Fig.~\ref{fig:k40-feshbach-resonance}, $a_s<0$
between $B_0$ and the zero crossing at $B_{\rm zero}=210.1\,\mathrm{G}$.
This interval corresponds to attractive interactions; for example,
$a_s(205\,\mathrm{G})=-306a_0$.

To construct the Wannier functions, we first consider a one-dimensional
lattice $V(x)=V_0\sin^2(k_Lx)$. With $v_0=V_0/E_r$ and
$E_r=\hbar^2k_L^2/(2m)$, the Hamiltonian in recoil units reads
\begin{equation}
    \frac{H_{{\rm single},1\mathrm{D}}}{E_r}
    =-\frac{1}{k_L^2}\frac{\partial^2}{\partial x^2}
    +v_0\sin^2(k_Lx).
    \label{eq:k40-single-1d}
\end{equation}
Bloch's theorem and the reciprocal vector $G=2k_L$ give
\begin{equation}
    \psi_{qn}(x)=e^{iqx}u_{nq}(x)
    =e^{iqx}\sum_{l\in\mathbb Z}c^n_{ql}e^{ilGx}.
    \label{eq:k40-bloch-expansion}
\end{equation}
In this basis the diagonal and nearest off-diagonal matrix elements are
\begin{equation}
    \frac{H_{ll'}}{E_r}
    =\left[\left(\frac{q}{k_L}+2l\right)^2+\frac{v_0}{2}\right]\delta_{ll'}
    -\frac{v_0}{4}\left(\delta_{l,l'+1}+\delta_{l,l'-1}\right).
    \label{eq:k40-plane-wave-matrix}
\end{equation}
After diagonalization, the band-$n$ Wannier function centered at $x_j$ is
\begin{equation}
    w_j^n(x)=\frac{1}{L}\sum_q e^{-iqx_j}\psi_{qn}(x).
    \label{eq:k40-wannier-transform}
\end{equation}
Here $L$ is the number of lattice sites, and the Bloch states are
normalized over one unit cell.
The calculation retains the lowest band and fixes the Bloch-state phase before
this sum.

For the scalar lattice Hamiltonian
$H_{{\rm single},3\mathrm{D}}=\mathbf p^2/(2m)+V_{\rm lat}(\mathbf r)$,
separability gives $H_{{\rm single},3\mathrm{D}}=H_x+H_y+H_z$ and
$w(\mathbf r)=w_xw_yw_z$.  The spin-conserving hopping amplitudes are obtained from the
Wannier matrix elements,

\begin{equation}
    -t_{ij}=\int d^3r\,w_i^*(\mathbf r)
    H_{{\rm single},3\mathrm{D}}w_j(\mathbf r).
    \label{eq:k40-hopping-definition}
\end{equation}
The lattice spacings are
$a_x=a_y=a_{xy}=\lambda_{xy}/2$ in the $xy$ plane
and $a_z=\lambda_z/2$ along the $z$ direction. For an $x$ hop the transverse integrals equal unity, whereas the $H_y,H_z$
terms contain $\int dx\,w_x^*(x)w_x(x-a_x)=0$.  Hence
\begin{equation}
    \begin{split}
    t_0&=-\int dx\,w_x^*(x)H_xw_x(x-a_{xy}),\\
    t_z&=-\int dz\,w_z^*(z)H_zw_z(z-a_z).
    \end{split}
    \label{eq:k40-spin-conserving-hopping}
\end{equation}
For the Raman fields in Eq.~\eqref{eq:k40-raman-fields}, the spin-flip
hopping along $x$ is obtained by evaluating the matrix element between
neighboring Wannier orbitals,
\begin{equation}
    t_{{\rm so},x}=-\int d^3r\,w_i^*(\mathbf r)
    \left(M_1+iM_2\right)w_{i+\hat x}(\mathbf r).
    \label{eq:k40-raman-matrix-element}
\end{equation}
The $M_2$ term vanishes because $|w_y|^2\sin(k_0y)$ is odd about a lattice
minimum; the remaining $M_1$ term gives
\begin{equation}
    \begin{split}
    t_{\rm so}={}&M_0\left|\int dx\,w_x(x)\sin(k_0x)w_x(x-a_{xy})\right|\\
    &\times\left|\int dy\,|w_y(y)|^2\cos(k_0y)\right|.
    \end{split}
    \label{eq:k40-spin-flip-hopping}
\end{equation}
To estimate the onsite interaction, we use the contact potential
\begin{equation}
    V_{\rm int}(\mathbf r_1,\mathbf r_2)
    =g\delta(\mathbf r_1-\mathbf r_2),\qquad
    g=\frac{4\pi\hbar^2a_s}{m}.
    \label{eq:k40-contact-potential}
\end{equation}
With $v_\nu=r_\nu/a_\nu$ and $\int dv_\nu|w_\nu|^2=1$, the physical Wannier
function is
\begin{equation}
    W(\mathbf r)=\frac{1}{\sqrt{a_xa_ya_z}}
    \prod_{\nu=x,y,z}w_\nu(v_\nu).
    \label{eq:k40-physical-wannier}
\end{equation}
Evaluating the onsite matrix element of the contact potential with
the Wannier function in Eq.~\eqref{eq:k40-physical-wannier} gives
\begin{equation}
    U_{\rm contact}=\frac{g}{a_xa_ya_z}
    \prod_{\nu=x,y,z}\int dv_\nu\,|w_\nu(v_\nu)|^4.
    \label{eq:k40-contact-interaction}
\end{equation}
In the special case of equal lattice spacings $a_x=a_y=a_z=a$,
Eq.~\eqref{eq:k40-contact-interaction} can be written in recoil units,
with $E_r=\hbar^2\pi^2/(2ma^2)$, as
\begin{equation}
    \frac{U_{\rm contact}}{E_r}
    =\frac{8a_s}{\pi a}
    \prod_{\nu=x,y,z}\int dv_\nu\,|w_\nu(v_\nu)|^4.
    \label{eq:k40-interaction-recoil}
\end{equation}
For $a_s<0$, $U_{\rm contact}<0$. To match the convention $H_{\rm int}=-U\sum_i\hat n_{i\uparrow}
\hat n_{i\downarrow}$ in the main text, we use
$U=-U_{\rm contact}>0$. Table~\ref{tab:k40-tight-binding-parameters} lists representative
experimental parameter sets for ${}^{40}\mathrm{K}$ at fixed lattice
wavelengths $\lambda_{xy}=764\,\mathrm{nm}$ and
$\lambda_z=1064\,\mathrm{nm}$, relating the optical and magnetic
controls to the tight-binding parameters.
For each in-plane lattice depth $V_{xy}$, the axial lattice
depth $V_z$ and Raman coupling strength $M_0$ are adjusted to obtain
comparable hopping amplitudes, $t_{\mathrm{so}}\simeq t_z\simeq t_0$.
The two magnetic fields listed for each optical configuration give
$U/t_0\simeq30$ and $20$, respectively, matching the interaction
regime considered in the main text.
More generally, varying the magnetic field continuously tunes $U/t_0$
across intermediate values while keeping the optical parameters fixed.

\clearpage

\bibliography{references}

\end{document}